\documentclass[aps,prd,%preprint,
notitlepage,11pt,showpacs,groupedaddress]{revtex4-1}

\usepackage{amsmath,amssymb,mathrsfs}
\usepackage{graphicx}
\begin{document}

%\preprint{}

\title{Hawking radiation from a collapsing dust sphere and its back~reaction at the event horizon ---Weak value approach---}


\author{Yuki Kanai}
%\thanks{}
\affiliation{Department of Physics, Tokyo Institute of Technology, Tokyo 152-8551, Japan}

\author{Akio Hosoya}
%\thanks{}
\affiliation{Department of Physics, Tokyo Institute of Technology, Tokyo 152-8551, Japan}


\date{\today}

\begin{abstract}
To see the back reaction of the Hawking radiation in a dynamical spacetime of the spherical gravitational collapse, 
we explicitly calculate the weak value of the energy-momentum tensor of the massless scalar field. 
The background geometry of a collapsing dust sphere is specified by using the Painlev\'e-Gullstrand coordinates, 
in which the time coordinate coincides with the proper time of a free-falling observer and the metric tensor is regular at the event horizon. 
The result is that in the remote future the weak value diverges at the event horizon. 
We argue that since the semi-classical approximation of the Einstein equation in the sense of the weak value breaks down there, 
the future geometry of the spacetime cannot be the Schwarzschild geometry. 
\end{abstract}

\pacs{03.65.Ta, 04.62.+v, 04.70.Dy}
%\keywords{}

\maketitle

\section{Introduction}%%
In the context of classical general relativity, black holes only absorb particles but do not emit. 
Absorption of particles or energy would make a black hole larger. 
Hawking mathematically showed that the area $A$ of the event horizon of each black hole does not decrease in time, $\delta A\geq 0$~\cite{H1972} in classical gravity. 
Combining the black hole physics and thermodynamics, 
Beckenstein suggested that a black hole has its intrinsic entropy proportional to its surface area~\cite{B1973}. 
This property of black holes is analogous to the second law of thermodynamics. 
Bardeen, Carter, and Hawking have shown that classical black holes with the mass $M$, the angular momentum $J$ and the electric charge $Q$ 
satisfy the four laws in analogy to thermodynamics~\cite{BCH1973}. 
The first law is that any two stationary axisymmetric black holes are related by $\delta M = (\kappa/8\pi) \delta A + \Omega_{\mathrm{H}}\delta J$, 
where $\kappa$ is the surface gravity and $\Omega_{\mathrm{H}}$ is the angular velocity of the black hole. 
The surface gravity $\kappa$ of a stationary black hole is analogous to temperature in the zeroth law; 
$\kappa$ is constant over the event horizon. 
The statement of the third law is that it is impossible to reduce $\kappa$ to zero by a physical process~\cite{W1994}. 

However, the temperature of a black hole is to be absolute zero since if it were not zero a classical black hole would emit radiation. 
This puzzle was resolved in quantum field theory in curved spacetime by Hawking 
proving that a black hole formed by gravitational collapse will create and radiate particles at a steady rate, 
which is by now called the Hawking radiation~\cite{H1974-75}. 
The radiated particles are considered to have the thermal distribution with a finite temperature $T=\kappa/2\pi$, which is called the Hawking temperature. 
This result gives the explicit form of black hole entropy, $S=A/4$. 
The area of the event horizon is likely to decrease due to the Hawking radiation. 
However, the sum of the quarter of the area $A$ and the entropy of the radiation never decreases as the generalized second law claims~\cite{B1974}. 
Thus, the Hawking radiation conforms with black hole thermodynamics. 

Each different choice of the coordinate system would generally give a different physical phenomenon such as the Hawking radiation in quantum field theories, 
since different time slices give different vacuum states. 
The spacetime geometry of the spherically symmetric (uncharged) black hole and the empty region outside a spherical body 
is given by the Schwarzschild solution. 
The Schwarzschild metric can be represented not only in the Schwarzschild coordinates but also in the Eddington-Finkelstein ones, the Kruskal-Szekeres ones, 
and so on~\cite{MTW1973}. 
Furthermore, it can also be expressed in the Painlev\'{e}-Gullstrand coordinates~\cite{P1921,G1922}. 
The last one is particularly useful for exploring the black hole physics, 
since the Painlev\'{e}-Gullstrand time coordinate coincides with the proper time of an observer freely falling from spatial infinity, 
while the Schwarzschild time coordinate coincides with the proper time of an observer being at rest at spatial infinity. 
The Painlev\'{e}-Gullstrand metric tensor is spatially flat and has an off-diagonal element so that it is regular at the event horizon. 
The hypersurface of a constant time traverses the event horizon to reach the central singularity. 
Exploiting the convenient properties of the  Painlev\'{e}-Gullstrand coordinates, 
we constructed a simple expression for the  gravitational collapse of a spherical dust star in the previous work~\cite{KSH2011}. 
We introduced a generalized form of the Painlev\'{e}-Gullstrand coordinates with the time coordinate being the proper time of a free-falling observer 
to describe both the interior matter region and the exterior empty region of a collapsing sphere in a single coordinate patch. 
It describes the inside region of the event horizon as well as the outside. 
In a semi-classical theory using the Painlev\'{e}-Gullstrand coordinates, the calculation of the Hawking radiation is performed, 
e.g., in~\cite{PW2000} as a tunneling process. 

In the present work, we are going to study \textit{the weak value} of the energy-momentum tensor in the Painlev\'{e}-Gullstrand coordinates. 
The weak value has been originally proposed by Aharonov and his collaborators~\cite{ABL1964,AAV1988,AR2005} in terms of the weak measurement. 
The weak value of an observable $A$ is defined by $<A_{\mathrm{w}}>:=\left<f\right|A\left|i\right>/\left<f|i\right>$, 
where $\left|i\right>$ and $\left<f\right|$ are an initial state and a final state, respectively. 
In general, the weak value is a complex number but can be measured by the weak measurement. 
Recently, many experiments have been performed to demonstrate the strange values~\cite{R2008,RLS2004}. 
Since then the weak value has been recognized as a fundamental concept of quantum mechanics~\cite{HS2010,HK2011}. 
Although it appeared in a different context, e.g., in quantum field theory, 
the explicit applications to black hole physics were pioneered by Englert and Spindel~\cite{ES2010}. 
They replaced the conventional vacuum expectation value by the weak value of the energy-momentum tensor 
to examine the back reaction of the Hawking radiation in the semi-classical approximation. 
Although the weak value was not directly calculated, they argued that it is not regular at the future event horizon 
and therefore the semi-classical approximation may break down there. 
Namely, the Hawking radiation changes the structure of the classical event horizon completely. 

In this paper, to see the back reaction of the Hawking radiation we directly calculate the weak value of the energy-momentum tensor in a dynamical spacetime of the spherical gravitational collapse 
and check its regularity at the future event horizon. 
By a new method in which the two kinds of the Bogoliubov transformation are considered 
because of the distinction between solutions of the Klein-Gordon equation in the empty region and in the matter region, 
we demonstrate the Hawking radiation of scalar particles in the Painlev\'{e}-Gullstrand coordinates. 
The Painlev\'{e}-Gullstrand coordinate system that is simple and well-behaved at the event horizon 
makes a matrix element of the weak value directly calculable. 
As a result, we will see that in the remote future the weak value of the energy-momentum tensor certainly diverges at the event horizon. 

Traditionally people in quantum field theory have studied the vacuum expectation value of the energy-momentum tensor 
to study the back reaction of the Hawking radiation 
and found that the incoming energy flux is negative, which is consistent with the decrease of the black hole mass~\cite{F1997}. 
We can reproduce this also in our weak value approach, 
since the vacuum expectation value is the weighted sum of the weak value with the weight being the probability to obtain the final state. 
Due to the large fluctuation it is obvious that the quantum state of matter cannot be semi-classical. 
However, we claim that the semi-classical gravity theory, in which the gravity field is treated as classical 
while the matter field is quantized and the energy-momentum tensor in the Einstein equation is replaced by its weak value, 
breaks down at the future event horizon. 

The organization of the present paper is as follows. 
We describe particle creation by the spherical dust collapse in the Painlev\'{e}-Gullstrand coordinates in section \ref{sec:pc}. 
To show the Hawking radiation and subsequent calculations, we specify the background geometry and the mode functions of the massless scalar field. 
The mode functions and the vacuum states reflect a physical feature of the coordinate system. 
In section \ref{sec:wv}, the weak value of the energy-momentum tensor of the field specified in section \ref{sec:pc} is calculated 
as the back reaction of the Hawking radiation, which appears in the Einstein equation in the semi-classical approximation. 
Taking the advantage of the nice properties of the Painlev\'{e}-Gullstrand coordinates, we directly calculate the weak value at the event horizon 
to see its dependence on the radial coordinate. 
Section \ref{sec:sd} is devoted to summary and discussions.

\section{Particle creation by a collapsing matter}\label{sec:pc}%%
\begin{figure}[h]
\begin{center}
\includegraphics[width=0.5\linewidth]{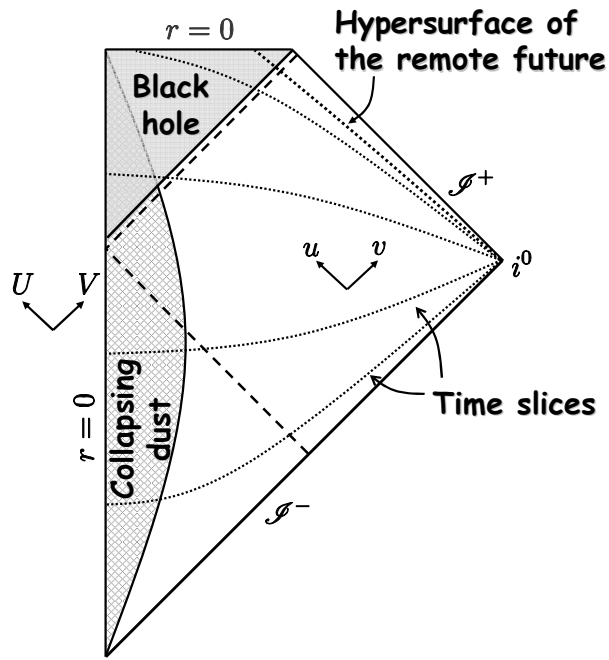}
\end{center}
\vspace{-0.5em}
\caption{The Penrose diagram for the emerging black hole. 
The in-vacuum state $\left|\mathrm{in}\right>$ is defined on the hypersurface of past infinity 
and the out-vacuum state $\left|\mathrm{out}\right>$ on the hypersurface of future infinity. 
}
\label{fig} 
\end{figure}

 In this section we demonstrate the occurrence of particle creation in the case of the spherical dust collapse employing the Painlev\'{e}-Gullstrand coordinates. 
The Painlev\'{e}-Gullstrand coordinates in the Schwarzschild spacetime specify a distinct time slice from 
that of the Schwarzschild or the Kruskal-Szekeres coordinates. 
The Painlev\'{e}-Gullstrand time coordinate has a physical meaning that it corresponds to the proper time of a free-falling observer 
not only in the empty Schwarzschild region but also in the collapsing dust region. 
In this coordinate system we show the particle creation in the black hole spacetime which emerges from the collapse. 
We ignore possible backscattering by the background geometry because it is not essential in the following discussion, 
and assume that if there exists particles which escape to the future null infinity 
they entirely come from the past null infinity propagating through the dust region (See FIG.~\ref{fig}). 

\subsection{Background geometry and mode functions}%
We assume that the geometry of the spherical collapsing dust spacetime is given by the metric form in the Painlev\'{e}-Gullstrand coordinates,
\begin{equation}
ds_{\pm}^2 = -dt^2+\left(dr+\sqrt{\frac{2m_{\pm}}{r}}dt\right)^2+r^2d\Omega^2, \label{metrics}
\end{equation}
where mass functions are $m_{+}=M(\mathrm{const.})$ in the exterior region of the dust sphere of mass $M$ 
and $m_{-}=2r^3/9t^2$ in the interior region~\cite{KSH2011}. 
In this case the dust sphere starts to collapse at rest in the remote past ($t \rightarrow -\infty$) 
and the radius of its surface is given by $r_{\mathrm{s}}(t)=\left(9M(-t)^2/2\right)^{1/3}$, $(-\infty<t<0)$. 
The radius of the sphere is $2M$ at time $t=-4M/3$ and is zero at $t=0$. 
The metric above is of $C^1$ class in the whole spacetime. 
The relation between the Schwarzschild time coordinate $t_{\mathrm{sch}}$ and the Painlev\'{e}-Gullstrand coordinates $(t, r)$ is 
\begin{equation}
dt_{\mathrm{sch}} = dt -\frac{\sqrt{2M/r}}{1-2M/r}dr, 
\end{equation}
and therefore, 
\begin{equation}
t_{\mathrm{sch}} = t -2\sqrt{2Mr}+2M\log{\left|\frac{\sqrt{r/2M}+1}{\sqrt{r/2M}-1}\right|}. 
\end{equation}

The line element of the exterior geometry is rewritten as 
\begin{eqnarray}
ds_{+}^2 = \left(1-\frac{2M}{r}\right)\left(-dt_{\mathrm{sch}}^2+dr_{\ast}^2\right)+r^2d\Omega^2 %\nonumber\\
         = -\left(1-\frac{2M}{r}\right)dudv+r^2d\Omega^2, 
\end{eqnarray}
where $r_{\ast}=r+2M\log{\left|r/2M-1\right|}$ is the tortoise coordinate. 
The retarded and advanced null coordinates are respectively represented by 
\begin{align}
&u = t_{\mathrm{sch}} -r_{\ast} = t -\xi(r), \\
&v = t_{\mathrm{sch}} +r_{\ast} = t +\eta(r), 
\end{align}
where 
\begin{eqnarray}
&&\xi(r) = r+2\sqrt{2Mr}+4M\log{\left|\sqrt{r/2M}-1\right|}, \\
&&\eta(r) = r-2\sqrt{2Mr}+4M\log{(\sqrt{r/2M}+1)}.
\end{eqnarray}
In the asymptotic region $r \rightarrow \infty$, the solution of the massless Klein-Gordon equation $\Box \Phi=0$ is the linear combination of 
\begin{equation}
\frac{1}{r}e^{-i\omega u}Y_{lm}(\theta,\phi) \quad\text{and}\quad  \frac{1}{r}e^{-i\omega v}Y_{lm}(\theta,\phi) 
\end{equation}
for the outgoing and incoming components, respectively. 
We define the in-modes $\phi_{\omega}^{\mathrm{in}}$ by the incoming solution having the positive frequency with respect to the Painlev\'{e}-Gullstrand time 
on the past null infinity $\mathscr{I}^{-}$, namely, 
\begin{equation}
\phi_{\omega}^{\mathrm{in}} \sim \frac{1}{\sqrt{4\pi\omega}r}e^{-i\omega v},\quad 
u \rightarrow -\infty, \label{ex-in}
\end{equation}
where the spherical harmonics $Y_{lm}(\theta,\phi)$ are suppressed (from now on we drop the angular part of modes). 
The modes $\phi_{\omega}^{\mathrm{in}}$ are normalized with respect to the Klein-Gordon inner product 
\begin{align}
\left(f_{\omega},f_{\omega'}\right) 
&= -i\int\!d\Sigma^{\mu}\sqrt{-g}\:  \left(f_{\omega}\partial_{\mu}f_{\omega'}^{\ast}-\partial_{\mu}f_{\omega}f_{\omega'}^{\ast}\right) \nonumber\\
&= -i\int\!dr r^2 f_{\omega}\left(\overset{\longleftrightarrow}{\partial_{t}}
-\sqrt{\frac{2M}{r}}\overset{\longleftrightarrow}{\partial_{r}}\right)f_{\omega'}^{\ast}, 
\end{align}
which now reads 
\begin{equation}
\left(\phi_{\omega}^{\mathrm{in}},\phi_{\omega'}^{\mathrm{in}}\right) 
= -i\int_{-\infty}^{+\infty}\!dv \left(r\phi_{\omega}^{\mathrm{in}}\right)\overset{\longleftrightarrow}{\partial_{v}}\left(r\phi_{\omega'}^{\mathrm{in}\ast}\right). 
\end{equation}
The field operator is expanded in terms of the in-modes as 
\begin{equation}
\Phi = \int\!d\omega \left(a_{\omega}^{\mathrm{in}}\phi_{\omega}^{\mathrm{in}}+a_{\omega}^{\mathrm{in}\dagger}\phi_{\omega}^{\mathrm{in}\ast}\right). 
\end{equation}
The in-vacuum state is defined as 
\begin{equation}
a_{\omega}^{\mathrm{in}}\left|\mathrm{in}\right> = 0 
\end{equation}
by the annihilation operators $a_{\omega}^{\mathrm{in}}$ satisfying the commutation relation, 
$[a_{\omega}^{\mathrm{in}}, a_{\omega'}^{\mathrm{in}\dagger}]=\delta(\omega-\omega')$.

On the other hand, we define the out-modes $\phi_{\omega}^{\mathrm{out}}$ as the outgoing solution on the future null infinity $\mathscr{I}^{+}$, i.e., 
\begin{equation}
\phi_{\omega}^{\mathrm{out}} \sim \frac{1}{\sqrt{4\pi\omega}r}e^{-i\omega u}, \quad
v \rightarrow +\infty, \label{ex-out}
\end{equation}
which is normalized with respect to the inner product 
\begin{equation}
\left(\phi_{\omega}^{\mathrm{out}},\phi_{\omega'}^{\mathrm{out}}\right) 
= i\int_{-\infty}^{+\infty}\!d\xi \left(r\phi_{\omega}^{\mathrm{out}}\right)\overset{\longleftrightarrow}{\partial_{\xi}}\left(r\phi_{\omega'}^{\mathrm{out}\ast}\right) 
\end{equation}
on the time slice of the remote future where its radial range is $2M<r<\infty$. 
The field can also be expanded as 
\begin{equation}
\Phi = \int\!d\omega \left(a_{\omega}^{\mathrm{out}}\phi_{\omega}^{\mathrm{out}}+a_{\omega}^{\mathrm{out}\dagger}\phi_{\omega}^{\mathrm{out}\ast}\right) \label{Phi+}
\end{equation}
and the relevant vacuum state is the out-vacuum, 
\begin{equation}
a_{\omega}^{\mathrm{out}}\left|\mathrm{out}\right> = 0. 
\end{equation}

The particles propagating into the matter region from $\mathscr{I}^{-}$ eventually reflect at the center of the dust sphere 
and go to the remote future. 
Focusing on the particles escaping through nearby the event horizon, 
it makes sense to take into account the intermediate modes on the interior matter region. 
The interior line element takes the form 
\begin{equation}
ds_{-}^2 = r_{\mathrm{s}}^2\left(-dT^2 +dR^2\right)+r^2d\Omega^2 = -r_{\mathrm{s}}^2dUdV+r^2d\Omega^2, 
\end{equation}
where the new time and radial coordinates are 
\begin{equation}
T = -\left(\frac{6(-t)}{M}\right)^{1/3},\quad 
R = \left(\frac{2}{9M(-t)^2}\right)^{1/3}r
\end{equation}
and then the null coordinates in the interior region are given by 
\begin{equation}
U = T -R,\quad V = T +R, 
\end{equation}
having the ranges $-\infty < U <-1$ and $-\infty < V < 1$. 
We denote the two independent exact solutions of the wave equation as 
\begin{eqnarray}
\psi_{\omega}^{\mathrm{out}} &=& \frac{1}{\sqrt{4\pi \omega}r}e^{-i\omega U}\left(1+\frac{1}{i\omega T}\right), \label{in-exactout}\\
\psi_{\omega}^{\mathrm{in}} &=& \frac{1}{\sqrt{4\pi \omega}r}e^{-i\omega V}\left(1+\frac{1}{i\omega T}\right), \label{in-exactin}
\end{eqnarray}
which are the outgoing and incoming modes in the interior region, respectively. 
We may impose the boundary condition that the mode functions should be regular at the origin $r=0$ so that 
the intermediate functions in the matter region appear only in the combination $\Psi_{\omega}=(\psi_{\omega}^{\mathrm{out}}-\psi_{\omega}^{\mathrm{in}})$. 

Massless particles propagate along null geodesics. 
Since the retarded and advanced null coordinates in the exterior geometry are continuously related to those defined in the interior 
at the surface of the dust sphere of radius $r_{\mathrm{s}}$, the two sets of coordinates have the functional relations 
\begin{align}
u = \frac{M}{6}\left[10+9U+U^3-24\log\left|\frac{-3-U}{2}\right|\right], \label{uU}\\
v = \frac{M}{6}\left[-10+9V+V^3+24\log\left|\frac{3-V}{2}\right|\right]. \label{vV}
\end{align}
The retarded null relation $(\ref{uU})$ diverges at the event horizon $U = -3$. 
Concerning the latest null ray that comes from $\mathscr{I}^{-}$ and bounces off at the center of the sphere to escape to $\mathscr{I}^{+}$, 
Eq.~$(\ref{uU})$ reduces to 
\begin{equation}
U(u) \sim U_0 -ae^{-\kappa u} \label{Uu}
\end{equation}
near the horizon $u \sim +\infty$ or $U \sim U_0 = -3$, 
where $\kappa = 1/4M$ and $a$ is a positive constant. 
On the other hand, Eq.~$(\ref{vV})$ reduces for the latest incoming ray to 
\begin{equation}
V(v) \sim b +cv, \label{Vv} 
\end{equation}
where $b$ is a certain real number and $c$ is a positive constant proportional to $\kappa$. 
Therefore, the outgoing modes in the interior region may be regarded as the outgoing modes in the remote future (in the exterior region) 
\begin{equation}
\psi_{\omega}^{\mathrm{out}} \sim \frac{1}{\sqrt{4\pi \omega}r}e^{-i\omega U(u)}, \quad
v \rightarrow +\infty \label{in-out}
\end{equation}
and the incoming modes as the ones in the remote past 
\begin{equation}
\psi_{\omega}^{\mathrm{in}} \sim \frac{1}{\sqrt{4\pi \omega}r}e^{-i\omega V(v)}, \quad
u \rightarrow -\infty. \label{in-in}
\end{equation}
In the interior region the intermediate mode solution appears only in the combination, 
$\Psi_{\omega}=(\psi_{\omega}^{\mathrm{out}}-\psi_{\omega}^{\mathrm{in}})$ by $(\ref{in-exactout})$ and $(\ref{in-exactin})$. 
However, in the remote past there exists only the incoming modes $\Psi_{\omega} \rightarrow -\psi_{\omega}^{\mathrm{in}}$ by $(\ref{in-in})$ 
and in the remote future only the outgoing modes $\Psi_{\omega} \rightarrow \psi_{\omega}^{\mathrm{out}}$ by $(\ref{in-out})$.

\subsection{Bogoliubov coefficients and Hawking radiation}%
We calculate the Bogoliubov coefficients which relate the in- and out-modes. 
The out-modes can be expanded in terms of the in-modes as 
\begin{equation}
\phi_{\omega}^{\mathrm{out}} 
= \int\!d\omega'' \left(\alpha_{\omega \omega''} \phi_{\omega''}^{\mathrm{in}} +\beta_{\omega \omega''} \phi_{\omega''}^{\mathrm{in}\ast}\right) \label{Beq}
\end{equation}
and also in terms of the intermediate modes as 
\begin{equation}
\phi_{\omega}^{\mathrm{out}} 
= \int\!d\omega' \left(A_{\omega \omega'} \Psi_{\omega'} +B_{\omega \omega'} \Psi_{\omega'}^{\ast}\right). \label{B-out}
\end{equation}
Here we have neglected the backscattering by the exterior Schwarzschild geometry 
and assumed that all the particles reaching $\mathscr{I}^{+}$ propagate through the interior region (See FIG.~\ref{fig}). 
Particularly in the remote future Eq.~$(\ref{B-out})$ reduces to 
\begin{equation}
\phi_{\omega}^{\mathrm{out}} 
= \int\!d\omega' \left(A_{\omega \omega'} \psi_{\omega'}^{\mathrm{out}} +B_{\omega \omega'} \psi_{\omega'}^{\mathrm{out}\ast}\right). 
\end{equation}
One can calculate the coefficients $A_{\omega \omega'} = \left(\phi_{\omega}^{\mathrm{out}},\psi_{\omega'}^{\mathrm{out}}\right)$ 
and $B_{\omega \omega'} = -\left(\phi_{\omega}^{\mathrm{out}},\psi_{\omega'}^{\mathrm{out}\ast}\right)$, 
using Eqs. $(\ref{ex-out})$ and $(\ref{in-out})$, to obtain 
\begin{equation}
\left\{\begin{array}{ll}
A_{\omega \omega'} \\
B_{\omega \omega'} 
\end{array}\right\} 
= \mp\frac{i\left(a\omega'\right)^{-i\omega/\kappa}}{2\pi \sqrt{\omega\omega'}} \exp{\left[\pm\left(\frac{\pi\omega}{2\kappa} + i\omega' U_0\right)\right]}
\Gamma\left(1+\frac{i\omega}{\kappa}\right), 
\end{equation}
where $\Gamma(z)$ is the gamma function. 

Comparing Eqs.~$(\ref{Beq})$ and $(\ref{B-out})$, we see that the two kinds of the coefficients have a relation 
\begin{equation}
\int\!d\omega' \left(A_{\omega \omega'} \Psi_{\omega'} +B_{\omega \omega'} \Psi_{\omega'}^{\ast}\right) 
= \int\!d\omega'' \left(\alpha_{\omega \omega''} \phi_{\omega''}^{\mathrm{in}} +\beta_{\omega \omega''} \phi_{\omega''}^{\mathrm{in}\ast}\right), \label{B-in}
\end{equation}
and on $\mathscr{I}^{-}$ it reduces to 
\begin{equation}
\int\!d\omega' \left(-A_{\omega \omega'} \psi_{\omega'}^{\mathrm{in}} -B_{\omega \omega'} \psi_{\omega'}^{\mathrm{in}\ast}\right) 
= \int\!d\omega'' \left(\alpha_{\omega \omega''} \phi_{\omega''}^{\mathrm{in}} +\beta_{\omega \omega''} \phi_{\omega''}^{\mathrm{in}\ast}\right). 
\end{equation}
The coefficients $\alpha_{\omega \omega''}$ and $\beta_{\omega \omega''}$ 
which relate the in-modes $\phi_{\omega''}^{\mathrm{in}}$ and the out-modes $\phi_{\omega}^{\mathrm{out}}$ can be obtained as 
\begin{equation}
\alpha_{\omega \omega''} 
= -\int_{0}^{\infty}\!d\omega' \left\{A_{\omega \omega'} \left(\psi_{\omega'}^{\mathrm{in}},\phi_{\omega''}^{\mathrm{in}}\right) 
+B_{\omega \omega'} \left(\psi_{\omega'}^{\mathrm{in}\ast},\phi_{\omega''}^{\mathrm{in}}\right)\right\} \label{B-a}\\
\end{equation}
and 
\begin{equation}
\beta_{\omega \omega''} 
= \int_{0}^{\infty}\!d\omega' \left\{A_{\omega \omega'} \left(\psi_{\omega'}^{\mathrm{in}},\phi_{\omega''}^{\mathrm{in}\ast}\right) 
+B_{\omega \omega'} \left(\psi_{\omega'}^{\mathrm{in}\ast},\phi_{\omega''}^{\mathrm{in}\ast}\right)\right\}. \label{B-b}
\end{equation}
Calculating first the inner product in Eqs.~$(\ref{B-a})$ and $(\ref{B-b})$, one obtains 
\begin{equation}
\left(\psi_{\omega'}^{\mathrm{in}},\phi_{\omega''}^{\mathrm{in}}\right) 
= \frac{ie^{-i\omega' b}}{4\pi\sqrt{\omega'\omega''}}\frac{\omega'+\omega''/c}{\omega'-\omega''/c+i\epsilon}, \quad
\left(\psi_{\omega'}^{\mathrm{in}\ast},\phi_{\omega''}^{\mathrm{in}}\right) 
= \frac{ie^{i\omega' b}}{4\pi\sqrt{\omega'\omega''}}\frac{\omega'-\omega''/c}{\omega'+\omega''/c-i\epsilon} 
\end{equation}
and 
\begin{equation}
\left(\psi_{\omega'}^{\mathrm{in}},\phi_{\omega''}^{\mathrm{in}\ast}\right) 
= \frac{ie^{-i\omega' b}}{4\pi\sqrt{\omega'\omega''}}\frac{\omega'-\omega''/c}{\omega'+\omega''/c+i\epsilon}, \quad
\left(\psi_{\omega'}^{\mathrm{in}\ast},\phi_{\omega''}^{\mathrm{in}\ast}\right) 
= \frac{ie^{i\omega' b}}{4\pi\sqrt{\omega'\omega''}}\frac{\omega'+\omega''/c}{\omega'-\omega''/c-i\epsilon}. 
\end{equation}
Finally the integration over $\omega'$ in Eqs.~$(\ref{B-a})$ and $(\ref{B-b})$ gives the Bogoliubov coefficients which relate the in- and out-modes, 
\begin{eqnarray}
\left\{\begin{array}{ll}
\alpha_{\omega \omega''} \\
\beta_{\omega \omega''} 
\end{array}\right\}
= \pm \frac{i\left(a\omega''/c\right)^{-i\omega/\kappa}}{2\pi\sqrt{\omega\omega''}} 
\exp\left[\pm \left(\frac{\pi\omega}{2\kappa} - i\omega''\frac{b - U_0}{c}\right)\right] \Gamma{\left(1 + \frac{i\omega}{\kappa}, \mp i\omega''\frac{b-U_0}{c}\right)},
\end{eqnarray}
where $\Gamma\left(z, t\right)$ is the incomplete gamma function. 

We express the out-modes in the remote future in the two different ways, i.e., by the in-modes in the remote past 
and by the intermediate modes in the matter region. 
The usual Bogoliubov coefficients are given by the integrals of the products of the Klein-Gordon inner products, 
coming from the relation of the two Bogoliubov transformations. 
Since the coefficients $\beta_{\omega \omega''}$ are nonzero, particle creation may occur in the situation considered in this section. 

\section{Weak value of energy-momentum tensor}\label{sec:wv}%%
Imagine an observer in a small box which freely falls and crosses the horizon. 
He is supposed to quantum mechanically measure the Hawking radiation. 
Suppose that the result is zero particle or vacuum defined with respect to the outgoing modes. 
The initial state is the vacuum with respect to the incoming modes. 
What we want to know is the weak value of the energy-momentum tensor, 
\begin{equation}
<T_{\mu\nu}>_{\mathrm{w}}:=\frac{\left<\mathrm{out}\right|T_{\mu\nu}\left|\mathrm{in}\right>}{\left<\mathrm{out}|\mathrm{in}\right>}, 
\end{equation}
where $\left|\mathrm{in}\right>$ and $\left<\mathrm{out}\right|$ are the in- and out-vacuum states of quantum matter fields. 

In this section, we calculate the weak value of the energy-momentum tensor of the scalar field using the results in the previous section. 
The back reaction of the Hawking radiation is given as the weak value of the energy-momentum tensor of quantum matter fields. 
Note that the weak value naturally appears on the right-hand side of the Einstein equation, 
\begin{equation}
R_{\mu\nu}-\frac{1}{2}g_{\mu\nu}R=8\pi<T_{\mu\nu}>_{\mathrm{w}}, 
\end{equation}
as justified, e.g., by the saddle point approximation in the path-integral quantization. 

In the remote future we have a reasonable form of the out-modes 
so that we can define the annihilation operators $a_{\omega}^{\mathrm{out}}$ 
and therefore the out-vacuum state $\left|\mathrm{out}\right>$ by $a_{\omega}^{\mathrm{out}} \left|\mathrm{out}\right> = 0$. 
Following the standard procedure of the Bogoliubov transformation, the in-vacuum state can be explicitly written in terms of the out-vacuum state as 
\begin{equation}
\left|\mathrm{in}\right> = \left<\mathrm{out}|\mathrm{in}\right> \exp\left[\frac{1}{2}\int\!d\omega d\omega' 
a_{\omega}^{\mathrm{out}\dagger}V_{\omega\omega'}a_{\omega'}^{\mathrm{out}\dagger}\right] \left|\mathrm{out}\right>, 
\end{equation}
where $V_{\omega\omega'}=-\left(\beta^{\ast}\alpha^{-1}\right)_{\omega\omega'}$. 
Then the weak value of the product of the fields is 
\begin{equation}
<\Phi(x)\Phi(x')>_{\mathrm{w}} = \int\!d\omega d\omega' \phi_{\omega}^{\mathrm{out}}(x) 
\left\{\delta_{\omega\omega'}\phi_{\omega'}^{\mathrm{out}\ast}(x') +V_{\omega\omega'}\phi_{\omega'}^{\mathrm{out}}(x')\right\}, \label{vev-PP}
\end{equation}
where $\delta_{\omega\omega'}$ is a short hand of the Dirac delta function $\delta(\omega-\omega')$. 
The out-modes are given by the outgoing solution $(\ref{ex-out})$ in the remote future. 
Their derivatives with respect to the time and radial coordinates are 
\begin{align}
&\partial_{t}{\phi_{\omega}^{\mathrm{out}}} = (-i\omega)\phi_{\omega}^{\mathrm{out}}, \label{deriv-t}\\ 
&\partial_{r}{\phi_{\omega}^{\mathrm{out}}} = \left(i\omega\partial_{r}{\xi}-r^{-1}\right)\phi_{\omega}^{\mathrm{out}}. \label{deriv-r}
\end{align}
Note that $\partial_{r}{\xi}=(1-\sqrt{2M/r})^{-1}\equiv\rho^{-1}(r)$ in (\ref{deriv-r}) diverges at the event horizon $r=2M$. 
Using Eq.~$(\ref{vev-PP})$, the weak value of the energy-momentum tensor 
$T_{\mu\nu}=\partial_{\mu}{\Phi}\partial_{\nu}{\Phi}-g_{\mu\nu}\partial^{\lambda}{\Phi}\partial_{\lambda}{\Phi}/2$ 
is given by 
\begin{equation}
<T_{\mu\nu}(x)>_{\mathrm{w}} = \int\!d\omega d\omega' \phi_{\omega}^{\mathrm{out}}(x) 
\left\{\delta_{\omega\omega'}\phi_{\omega'}^{\mathrm{out}\ast}(x)F_{\mu\nu}(x;\omega,\omega') 
- V_{\omega\omega'}\phi_{\omega'}^{\mathrm{out}}(x)G_{\mu\nu}(x;\omega,\omega')\right\}. \label{vev-st}
\end{equation}
Each component of $F_{\mu\nu}$ and $G_{\mu\nu}$ contains the derivatives such as $(\ref{deriv-t})$ and $(\ref{deriv-r})$. 
Their behaviors near the event horizon are 
\begin{align}
&F_{rr},\:G_{rr} = \omega\omega'\rho^{-2} +O(\rho^{-1}), \\
&F_{tr},\:G_{tr} = -\omega\omega'\rho^{-1} +O(\rho^{0}), \\
&F_{tt},\:G_{tt},\:\text{and the other components} = O(\rho^{0}). 
\end{align}
The divergent behavior is a direct consequence of the radial derivative $(\ref{deriv-r})$ of the outgoing solution, 
while $<\partial^{\lambda}{\Phi}\partial_{\lambda}{\Phi}>_{\mathrm{w}}$ is regular at the horizon. 
The leading behavior of the scalar $<T_{\mu\nu}u^{\mu}u^{\nu}>_{\mathrm{w}}$ is $\rho^{-2}$, 
where $u^{\mu}=(u^{t},u^{r})=(1,-\sqrt{2M/r})$ is the four-velocity of a radially free-falling observer. 

 Therefore, we conclude that the weak value of the energy-momentum tensor diverges at the future event horizon.

\section{Summary and discussions}\label{sec:sd}%%
We have demonstrated the Hawking radiation of scalar particles in a dynamical spacetime of the spherical gravitational collapse 
and then calculated the weak value of the energy-momentum tensor to see the back reaction at the event horizon of the emerging black hole
using the Painlev\'{e}-Gullstrand coordinates in which the metric tensor of the background geometry is regular at the event horizon. 
It is shown that some components of the weak value of the energy-momentum tensor are not regular at the future event horizon 
due to the dependence of the mode functions on the radial coordinate. 
Therefore, the semi-classical approximation of the Einstein equation breaks down. 

In our paper we have mainly concerned with the weak value of the energy momentum tensor, 
while people normally studied the vacuum expectation value in the semi-classical approximation. 
We think the vacuum expectation value is not enough to see the behavior of quantum back reaction. 
For example, the semi-classical approximation breaks down, if the weak values for a particular final state is divergent there, 
while the vacuum expectation value, the average of the weak values, is finite there. 
Note the relation of the weak value $<T_{\mu\nu}>_{\mathrm{w}}:=\left<f\right|T_{\mu\nu}\left|i\right>/\left<f|i\right>$ 
with the expectation value $\left<i\right|T_{\mu\nu}\left|i\right>$: 
\begin{equation} 
\left<i\right|T_{\mu\nu}\left|i\right>=\sum_{f}\left|\left<f|i\right>\right|^2<T_{\mu\nu}>_{\mathrm{w}}.
\end{equation}
Actually many calculations have shown that the vacuum expectation value is finite after regularization. 
We think that the divergences are cancelled out in the average. 

How can we physically interpret this? 
Consider an observer who is located far away from the star before its collapse. 
He will expect that a black hole will emerge and the Hawking radiation should occur. 
However, if someone detected a certain number of the Hawking particles or nothing at all, 
he should expect the resultant spacetime is far from the original Schwarzschild black hole 
because the conditioned value of the energy-momentum tensor is divergent at the future event horizon. 
The outcome spacetime is observer dependent. 

Englert and Spindel calculated the weak value in a clever but indirect way, e.g., via the trace anomaly. 
However, this obscures the physical origin of the divergence of the weak value. 
In our method, as mentioned above, it is obvious that the out-modes defined on the remote future give rise to the divergence at the event horizon. 
The out-mode function or rather the outgoing solution of the scalar field equation in the Schwarzschild spacetime 
breaks the semi-classical approximation. 
Hence we argue that the future geometry of the spacetime cannot be the Schwarzschild geometry. 

Several possible situations after a black hole evaporation have been considered by many people but with no consensus. 
Here we claim that the black hole formed by collapse never forms; 
the existence of the event horizon is inconsistent with the semi-classical Einstein equation. 
Even the lowest-order back reaction of the Hawking radiation breaks the structure of the event horizon critically. 

The Hawking radiation from black holes causes information loss paradox. 
Let us consider the situation in which a black bole is formed by a pure state. 
As the black hole emits thermal radiation and evaporates completely, the subsequent state is a mixed state. 
The transition from a pure state to a mixed state means a non-unitary evolution, which conflicts with quantum mechanics. 
Moreover, the existence of the event horizon implies the existence of the domain of out-of-communication with outside of the horizon. 
The event horizon leads to the non-unitary evolution by itself. 
If one believes that the process of the evaporation of black holes is unitary, 
the event horizon seems to be never formed. 

We would like to briefly comment on possible effects on the back reaction of the Hawking radiation 
from the Planckian physics~\cite{STU1993}. 
We have circumvented it in our modest claim that the Schwarzschild spacetime is not consistent 
with the semi-classical approximation. 
More precisely, the weak value of the energy-momentum tensor is described by the Painlev\'{e}-Gullstrand coordinates 
which is regular at the future event horizon. 
Even if one tried the weak measurement there, this would not require energy of Planck scale.

\begin{acknowledgments}
We would like to thank Dr. M. Hotta for his useful comments. 
The authors are supported by Global Center of Excellence Program ``Nanoscience and Quantum Physics" at Tokyo Institute of Technology. 
\end{acknowledgments}


\end{document}